\documentclass{bmcart}

\usepackage{amsthm,amsmath}
\usepackage[utf8]{inputenc} 
\usepackage{hyperref}

\usepackage{graphicx}
\usepackage{algorithm,algorithmic,multicol}
\usepackage{bm}
\usepackage{ulem}

\usepackage{bbding}
\usepackage{pifont}
\usepackage{wasysym}
\usepackage{amssymb}

\startlocaldefs
\endlocaldefs

\begin{document}

\begin{frontmatter}

\begin{fmbox}
\dochead{Research}

\title{
Bayesian Sequential Stacking Algorithm for Concurrently Designing Molecules and Synthetic Reaction Networks}

\author[addressref={ism},email={qiz@ism.ac.jp}] {\fnm{Qi} \snm{Zhang}}
\author[addressref={ism},email={liu.chang@ism.ac.jp}] {\fnm{Chang} \snm{Liu}}
\author[addressref={ism,sokendai},email={stewu@ism.ac.jp}] {\fnm{Stephen} \snm{Wu}}
\author[addressref={ism,sokendai,nims},corref={ism},email={yoshidar@ism.ac.jp}] {\fnm{Ryo} \snm{Yoshida}}

\address[id=ism]{%
  \orgname{Research Organization of Information and Systems, The Institute of Statistical Mathematics},
  \postcode{190-8562}
  \city{Tachikawa, Tokyo},
  \cny{Japan}
}

\address[id=sokendai]{%
  \orgname{The Graduate University for Advanced Studies, SOKENDAI},
  \postcode{190-8562}
  \city{Tachikawa, Tokyo},
  \cny{Japan}
}

\address[id=nims]{%
  \orgname{National Institute for Materials Science},
  \postcode{305-0047}
  \city{Tsukuba, Ibaraki},
  \cny{Japan}
}

\begin{artnotes}
\end{artnotes}

\end{fmbox}

\begin{abstractbox}

\begin{abstract} 
In the last few years, de novo molecular design using machine learning has made great technical progress but its practical deployment has not been as successful. This is mostly owing to the cost and technical difficulty of synthesizing such computationally designed molecules. To overcome such barriers, various methods for synthetic route design using deep neural networks have been studied intensively in recent years. However, little progress has been made in designing molecules and their synthetic routes simultaneously. Here, we formulate the problem of simultaneously designing molecules with the desired set of properties and their synthetic routes within the framework of Bayesian inference. The design variables consist of a set of reactants in a reaction network and its network topology. The design space is extremely large because it consists of all combinations of purchasable reactants, often in the order of millions or more. In addition, the designed reaction networks can adopt any topology beyond simple multistep linear reaction routes. To solve this hard combinatorial problem, we present a powerful sequential Monte Carlo algorithm that recursively designs a synthetic reaction network by sequentially building up single-step reactions. In a case study of designing drug-like molecules based on commercially available compounds, compared with heuristic combinatorial search methods, the proposed method shows overwhelming performance in terms of computational efficiency and coverage and novelty with respect to existing compounds.

\end{abstract}


\end{abstractbox}

\end{frontmatter}

\section*{Introduction}
In recent years, machine learning-based molecular design has made great technological advances and has achieved remarkable outcomes in drug discovery and materials science \cite{ikebata2017bayesian, wu2019machine, gomez2018automatic, segler2018generating, sanchez2018inverse, ramprasad2017machine}. The design objective here is to identify the chemical structure of a new molecule with a given set of desired properties. To do so, a machine learning model that forwardly predicts the physicochemical properties of any given chemical structure is first constructed and then its inverse mapping is found to determine the design of the chemical structure that exhibits the desired properties in the backward direction. The former is often referred to as quantitative structure--property relationship (QSPR) analysis \cite{roy2015understanding}, and the latter is called inverse structure--property relationship (inverse-QSPR) analysis \cite{miyao2016inverse}. Usually, the inverse problem is solved using heuristic search techniques such as a genetic algorithm. A molecular generator is then used to sequentially modify a candidate molecule such that the resulting predicted properties fall into the region of the desired properties. The generative model plays an important role in this process. Traditionally, structural modifications have been performed by conducting stochastic recombination with a predefined set of molecular fragments or random atomic substitution. By utilizing fragments of existing molecules as building blocks, we can restrict the degrees of freedom in the resulting chemical structures and narrow down the search space to enhance the synthesizability of virtually created molecules. However, excessive narrowing of the search space may reduce the novelty of the structures created. In order to overcome this limitation, increasing attention has been paid, since around 2017--2018 in particular, to the development of molecule generative models that rely on advanced machine learning techniques \cite{ikebata2017bayesian, yang2017chemts, assouel2018defactor,dai2018syntax, gomez2018automatic, jin2018junction, kadurin2017cornucopia, kajino2019molecular, li2018learning, seff2019discrete, segler2018generating, simm2019generative, simm2020reinforcement, you2018graph}. With a training set of compounds synthesized thus far, a generative model for molecular graphs is constructed to mimic the rule of frequently appearing chemical fragments and bonding patterns in the training molecules. Using such a model, we can freely scan the vast chemical space to identify innovative hypothetical molecules.

As described above, machine learning-based molecular design has made great technical progress over the past few years. However, its practical deployment has not progressed as much as expected. One reason for this is owing to the difficulty in determining the synthetic routes to such designed molecules. Concurrently , with the growing attention to molecular design, significant progress has been made in computational methods for synthetic route design \cite{schwaller2019molecular, jin_predicting_2017, liu2017retrosynthetic, lin2020automatic, zheng2019predicting}. Similar to the molecular design task, the general workflow for synthetic route design consists of forward and inverse problems. The goal of the forward problem is to derive a model that predicts the chemical structure of a synthetic product for a given set of reactant molecules. In contrast, in the inverse problem, inverse mapping of the forward model is explored to identify a set of reactants that produces a given desired product. Recent developments in deep learning technologies have significantly improved the accuracy of predicting reaction outcomes in organic synthesis. For example, if the structures of the reactants and products are treated as graphs, the reaction prediction task can be formulated as a graph transformation problem \cite{jin_predicting_2017}. Under the string representation of reactants and products according to the simplified molecular input line entry system (SMILES) chemical language \cite{weininger1988smiles}, deep neural networks for sequence-to-sequence translation can be utilized to predict the SMILES string of a product from input reactants. For example, it was reported that the reaction prediction using Transformer, a well-known encoder-decoder architecture for machine translation, could successfully predict the chemical structures of synthetic products with over ${90\%}$ accuracy \cite{schwaller2019molecular}. Furthermore, several methods have been developed to solve the inverse problem of such forward reaction prediction models. The objective is to identify a set of promising reactants from a list of commercially available compounds that can synthesize a desired product \cite{guo2020bayesian, bradshaw2020barking, gottipati2020learning}.

More recently, several attempts have been made to simultaneously design desired functional molecules and their synthetic routes under a unified methodological framework. For a given reaction prediction model, a virtual library of candidate molecules can be created by which a set of reactants is given to the model to produce its synthetic product. In addition, the properties or a certain score of a candidate molecule can be calculated using machine learning models or an arbitrary reward function. By obtaining the inverse mapping of such a cascade model, which defines the composite mapping from any given reactant set to a product and from the product to its physicochemical properties, the products exhibiting desired properties and their reactant sets can be predicted simultaneously. The Molecule Chef algorithm proposed by Bradshaw et al. \cite{bradshaw2019model} uses a cascaded forward model that connects a deep generative model of reactant molecules (Molecular Transformer) for the reaction prediction \cite{schwaller2019molecular}, and a property prediction model for the created virtual products. Gottipati et al. \cite{gottipati2020learning} formulated the inverse problem of such a cascade model as a combinatorial optimization problem over a set of commercially available compounds, and proposed a reinforcement learning algorithm to identify the promising combination of reactant molecules. However, there are still many unsolved technical problems in these existing methods. In addition, their predictive performance and systematic evaluation from the viewpoint of organic chemistry are insufficient, and they have not yet reached the level required for practical adoption. 

In this paper, we formulate the task of simultaneous design of molecules and synthetic routes as a general statistical problem in which a forward model is defined as a cascade of reaction prediction models and property prediction models, as in the previous work of Gottipati et al. \cite{gottipati2020learning}. In the inverse problem, we seek a set of reactants and a network structure of synthetic reactions such that the resulting products reach a predefined region of desired properties. The overall search space for the reactant sets is spanned by all possible combinations of given commercially available compounds. If the number of commercial compounds is of the order $O(10^{6})$ and the number of reactants involved in a synthetic route is 10, the size of the search space will reach $O(10^{60})$. The structure of the reaction network is a design variable as well, involving the number of reaction steps and the network topology. In conventional computer-aided synthetic route design, only a single-chain linear reaction is considered as a design target, but branching routes, the width and depth of networks, and the number of leaf nodes should also be included in the design variable. Therefore, it is necessary to solve the intractable combinatorial problem defined over an extremely vast search space. 

\begin{figure}[h!]
\includegraphics[width=0.95\linewidth]{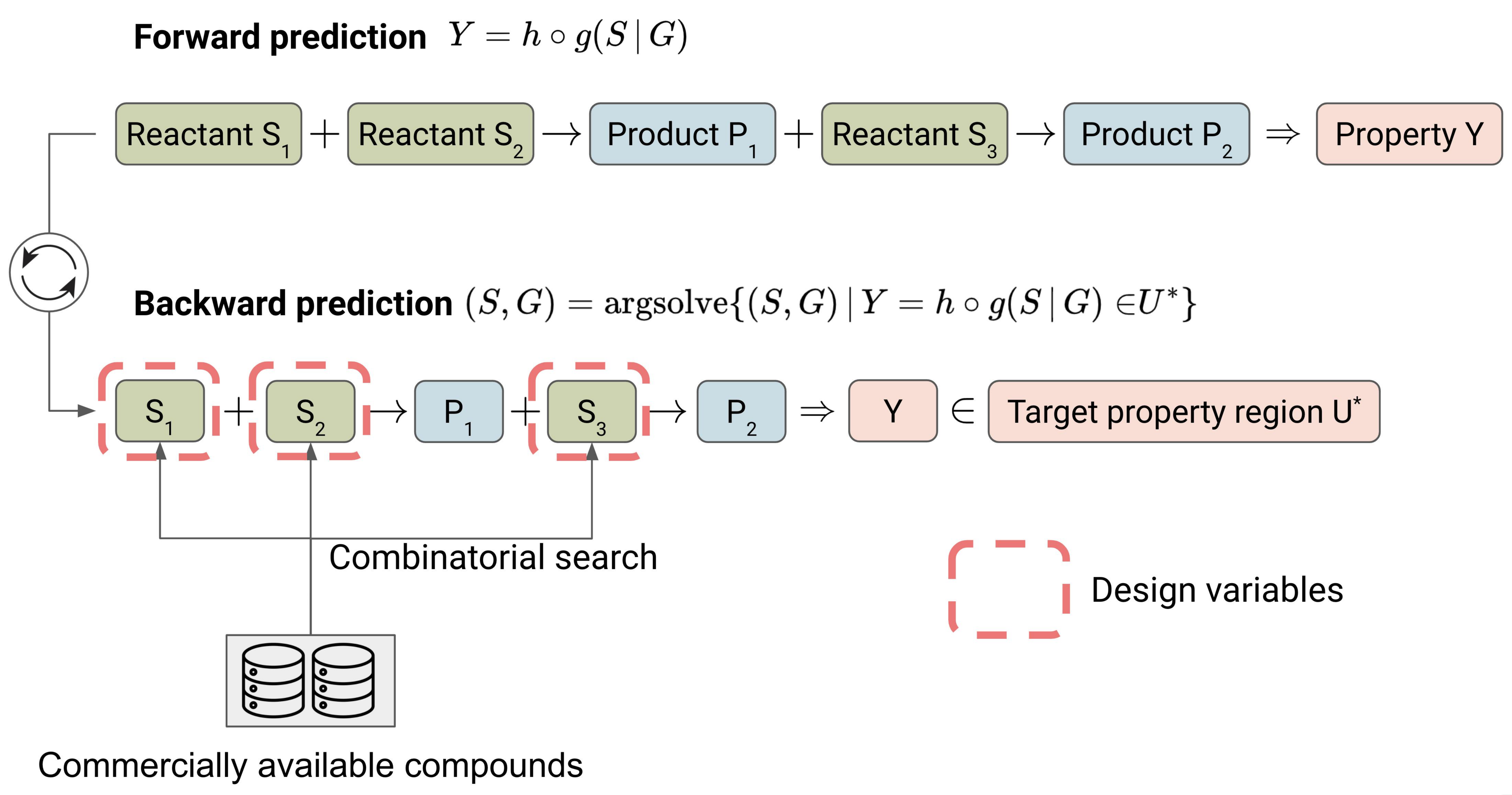}
\caption{
Workflow of the concurrent design of molecules and their synthetic reaction networks. The forward-prediction model defines the composite mapping $h \circ g$ from a given reactant set $S$ to its synthetic product $P$ conditioning on a network $G$ via a reaction function $g$, and from product $P$ to its properties $Y$ via a prediction model $h$. The desirable products and their synthetic reactions are concurrently designed by inversely exploring $h$ and $S$ such that the resulting properties meet the design objective $Y \in U^*$.
}
\label{fig:summary}
\end{figure}

Here, we present an efficient algorithmic technique and implementation to solve this hard problem. Specifically, we present a sequential Monte Carlo method based on a recurrent network search algorithm to simultaneously identify a reaction network, its constituent reactant sets, and the final products that satisfy the arbitrary target properties. An important practical requirement considered here is the maintenance of diversity in the designed molecules and synthetic routes. There are always errors in the properties and synthetic routes predicted by statistical models. Therefore, the optimal solution in a model is not optimal in practice. Our method aims to exhaustively identify a wide variety of promising candidates and to present various scenarios to domain experts. The final decision is left to the expert. A Python library called Seq-Stack-Reaction has been made available on GitHub\cite{ssr}, which allows us to plug-in any forward model into the Bayesian design workflow (Figure \ref{fig:summary}). 

\section*{Methods}
\subsection*{Forward model: synthetic reaction}
Suppose that a single-step reaction is given as
\begin{align}
    S_{1} + S_{2} \rightarrow P.
\end{align}
$S_1$ and $S_2$ denote two reactants, and $P$ denotes the product, which is assumed to be a singleton as byproducts are ignored here. In this study, we considered only synthetic reactions with two reactants, but the proposed method can be generalized to handle any number of reactants. In addition, solvents and reagents can also be incorporated on demand. Consider that the single-step reaction is modeled by a function $r$ as
\begin{align}
    P & = r(S),
\end{align}
where $S = \{S_1, S_2\}$. Function $r$ represents the change in the chemical structure from $S$ to $P$. Note that $r$ is a set function that is invariant to the exchange of $S_1$ and $S_2$. As described later, the single-step reaction prediction model was modeled using Molecular Transformer.

\begin{figure}[h!]
\includegraphics[width=0.95\linewidth]{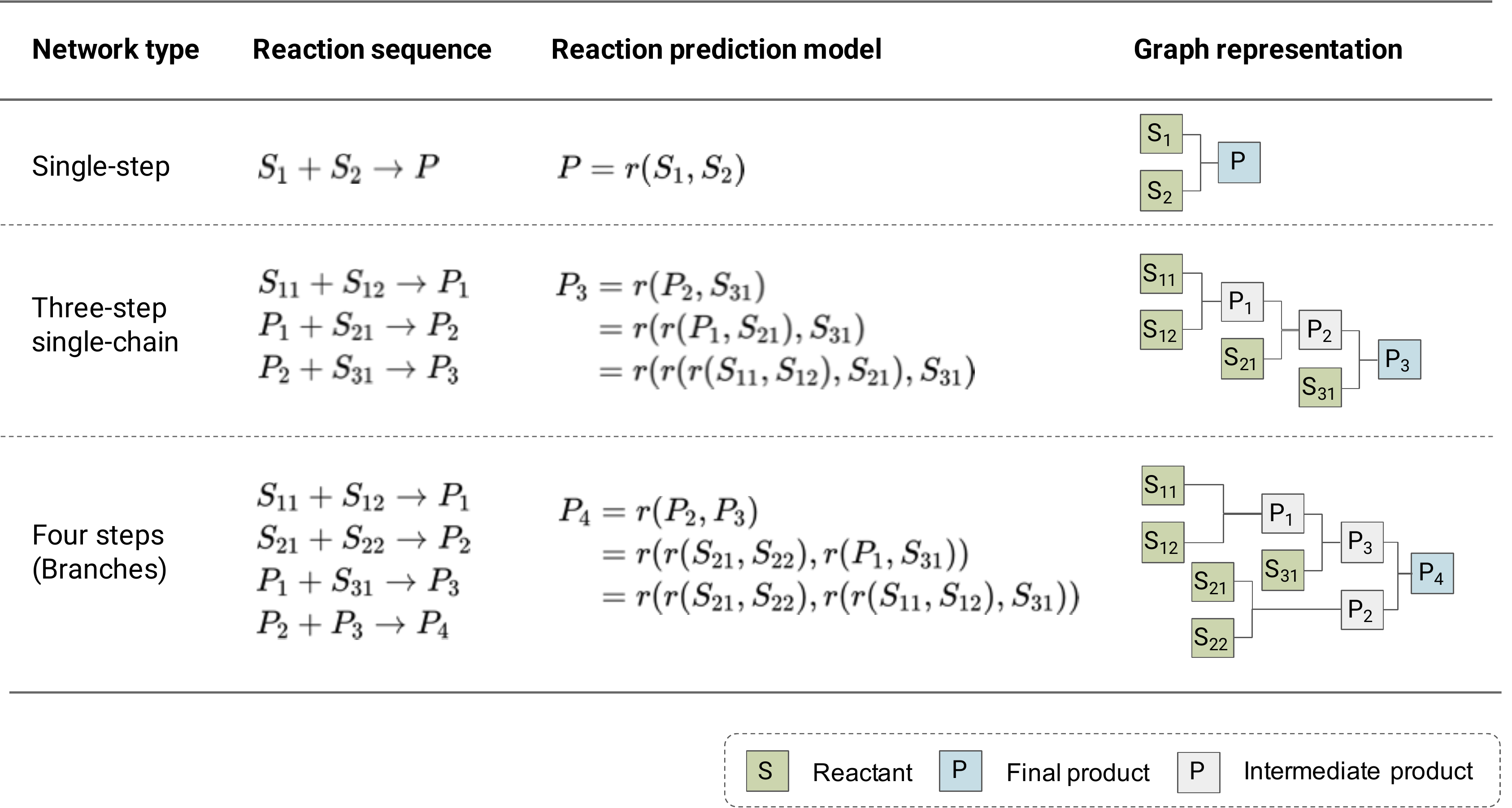}
\caption{Examples of synthetic reaction networks.}
\label{fig:example}
\end{figure}

Arbitrary reaction networks can be modeled by combining the single-step model. For example, consider a three-step single-chain reaction as 
\begin{equation}
  \begin{split}
    & \text{Step 1}:  S_{11} + S_{12}  \rightarrow P_{1}\\
    & \text{Step 2}:  P_{1} + S_{21}  \rightarrow P_{2}\\
    & \text{Step 3}:  P_{2} + S_{31}  \rightarrow P_{3} \ (=: P).
    \end{split}
\end{equation}
In the first step, two reactants, $S_{11}$ and $S_{12}$, produce the intermediate product $P_{1}$, followed by the second step, which produces the second intermediate product $P_{2}$ by reacting $P_1$ and a newly selected reactant $S_{21}$.
In the third step, the intermediate product $P_{2}$ and reactant $S_{31}$ react to produce final product $P := P_{3}$. This reaction cascade can be expressed using the single-step reaction model, as follows:
\begin{equation}
  \begin{split}
    P & = r(P_{2}, S_{31})\\
    & = r(r(P_{1}, S_{21}), S_{31})\\
    & = r(r(r(S_{11}, S_{12}), S_{21}), S_{31})\\
    & =: g(S|G).
    \end{split}
\end{equation}
The final product $P = P_3$ is described as a function $g(\cdot|G)$ of the four purchasable reactants $S=\{S_{11}, S_{12}, S_{21}, S_{31}\}$, where all the intermediate reaction states are discarded. The structure of the reaction network is represented by the synthetic graph or network $G$. The graph forms a rooted tree in which the leaf nodes consist of the four reactants in $S$ and the root node is given by the final product. Every node, except the leaf nodes, has two children. Without loss of generality, any synthetic reaction network, beyond single-chain reactions, can be described as $P = g(S|G)$ and retains the graph properties. Several examples are shown in Figure \ref{fig:example}.

\subsubsection*{Forward model: property prediction models and scoring function}
Once the final product $P$ is generated as $P = g(S|G)$ with any given $S$, we estimate its properties $Y$ using the prediction model $Y = h(P)$ as
\begin{equation}
  \begin{split}
    Y & = h(P) \\
      & = h \circ g(S|G) \\
      & =: f(S|G).
  \end{split}
    \label{eq:prop}
\end{equation}
As the product can be represented by the deterministic function $g(S|G)$ of $S$, $Y$ can be expressed by a function of $S$ as $Y = f(S|G)$. Here, $Y$ can be a vector of one or more properties. Function $h$ can be specified arbitrary; for example, a machine learning property prediction model or a scoring function such as the quantitative estimate of drug-likeness (QED) score \cite{bickerton2012quantifying}.

The objective of molecular design is to identify a reactant set $S$ with the resulting $P$ exhibiting a set of desired properties $Y^{*}$ with respect to the given forward model $Y = f(S|G)$. For the molecular design task, it is necessary to define a measure of the discrepancy $d(Y, Y^*)$ between the predicted properties and the target. A typical example of a discrepancy measure is the Euclidean distance:
\begin{align}
    d(Y, Y^{*}) = \|Y - Y^{*}\|^{2}.
\end{align}
Additionally, various measures can be defined depending on the type of task. For example, if the target property is given as a region $U^*$, we can use the following 0-1 loss:
\begin{align}
    d(Y, U^{*}) =
  \begin{cases}
    0  & \quad Y \in U^{*}\\
    1  & \quad \text{otherwise}
  \end{cases}
  .
\end{align}
Hereafter, we consider $d(Y, U^{*})$, but we can use any type of discrepancy without loss of generality.
Alternatively, in the case of a score-type monotonic measure such as QED, $d$ can be defined, for example, as
\begin{align}
    d(Y) = -\text{QED}(Y).
\end{align}
Furthermore, in the task of retrosynthetic prediction, where a target product $P^*$ is given as the design purpose, we can use, for example, the 0-1 loss between $P$ and $P^*$ as follows:
\begin{align}
    d(P, P^{*}) =
  \begin{cases}
    0  & \quad P = P^{*}\\
    1  & \quad \text{otherwise}
  \end{cases}
  .
\end{align}
This is equivalent to setting $h(P)=1$ and $Y=P$ in the property prediction model, Eq. \ref{eq:prop}. 

\subsection*{Bayesian inverse problem}
Herein, we describe the task of designing molecules with their synthetic routes. Suppose that a collection of $N$ commercially available compounds is given by
\begin{align}
    \mathcal{B} = \{S_1, ..., S_N\}.
\end{align}
A reactant set $S$ that forms the designed $G$ should be selected from $\mathcal{B}$. Here, by $\mathcal{P}_k(\mathcal{B})$, we denote the set of all $k$ combinations from $\mathcal{B}$. Then,
the support $\mathcal{P}(\mathcal{B})$ of $S$ is expressed as
\begin{eqnarray}
\mathcal{P}(\mathcal{B}) = \mathcal{P}_1(\mathcal{B}) + \cdots + \mathcal{P}_K(\mathcal{B}),
\label{eq:space}
\end{eqnarray}
where the maximum number of reactants that can be selected is constrained to $K$.

The design variable is denoted by a tuple $x = \{S, G\}$. Here, we write the model in Eq. \ref{eq:prop} as $Y(x)=f(x)=f(S|G)$ in order to explicitly express that the predicted property $Y$ is a deterministic function of the reactant $S$ and the reaction network $G$. As in our previous studies \cite{ikebata2017bayesian, guo2020bayesian}, molecular design based on Bayesian inference is performed based on the target distribution $\pi(x)$, which is defined on $\mathcal{P}(\mathcal{B})$:
\begin{align}
\begin{split}
    \pi(x) & = p(x | Y \in U^{*})\\
    & \propto p(x, Y \in U^{*})\\
    & = p(Y \in U^{*} | x) p(x)\\
    & = \frac{1}{Z}\exp\left(-\frac{1}{\sigma}d(Y(x), U^{*})\right) \cdot p(x).
    \end{split}
\end{align}
According to Bayes' law, the posterior distribution $\pi(x) = p(x|Y \in U^{*})$ is proportional to the joint distribution $p(x, Y \in U^{*})$, which consists of the product of the likelihood $p(Y \in U^{*} | x)$ and prior distribution $p(x)$. The forward model forms the joint probability distribution, which is modeled by the Gibbs distribution $p(Y \in U^{*} | x) = \frac{1}{Z}\exp\left(-\frac{1}{\sigma}d(Y(x), \textbf{U}^{*})\right)$, with temperature parameter $\sigma > 0$. The normalizing constant $Z$ is the canonical partition function:
\begin{align}
\begin{split}
    Z & = \sum_{Y \in U^{*}, Y \notin U^{*}} \exp\left(-\frac{1}{\sigma}d(Y(x), U^{*})\right), \\.
    & = \exp\left(-\frac{1}{\sigma}\times 0\right) + \exp\left(-\frac{1}{\sigma}\times 1\right)\\
    & = 1 + \exp\left(-\frac{1}{\sigma}\right)
    \end{split}
\end{align}
The prior distribution $p(x)$ is used to narrow down the broad solution space based on prior knowledge. For example, the prior can be modeled as follows:
\begin{align}
    p(x) \propto I(S \subset \mathcal{P}(\mathcal{B}))\cdot \exp\left(-\frac{|S|}{\tau}\right) \cdot \exp\left(-\frac{|G|}{\tau^{\prime}}\right).
\end{align}
The indicator function $I(\cdot)$ takes the value one or zero depending on whether the designed reactant set $S$ belongs to a subset of purchasable compounds. The second and third terms on the right-hand side penalize the increasing number of reactants ($|S|$) and the size of the designed network ($|G|$), where $\tau$ and $\tau^{\prime}$ determine the magnitude of penalties.

By identifying $x$ with a sufficiently high posterior probability, we predict product $P$ and its synthetic reaction route $\{S, G\}$ that satisfy design objective $U^*$. However, computational difficulty here arises from the extremely large search space. The search space is composed of all combinations of reactants that are purchasable. The number of candidate reactants is typically of the order $O(10^6)$, resulting in the cardinality of the solution space burgeoning to $O(10^{6\times k})$ when $k$ reactants are involved in the synthetic route planning. In addition, the network topology to be explored further increases the size of the search space.

\subsection*{Sequential Monte Carlo in general}

For the main building block of the proposed method, we employed a sequential Monte Carlo (SMC) algorithm \cite{del2006sequential} to draw a promising sample set of $x = \{S, G\}$ from the target $\pi(x)$. As mentioned above, because the target distribution is defined on a large combinatorial space, an ordinary SMC cannot approximate it adequately. In particular, it is difficult to obtain a diverse set of highly probable molecules with their reaction routes using such ordinary methods, which will be demonstrated later. The proposed method, shown in the next subsection, was developed as an extension of SMC to overcome the difficulty of combinatorial complexity. To clarify the design concept of the proposed method, we briefly describe the general SMC method here.

In conventional SMC, we define an augmented target distribution $\pi_A(\bm{x})$ using $T$ auxiliary distributions, $\pi_t(x_t)$ $(t=1, \ldots, T)$, and an arbitrarily chosen initial distribution $\pi_0(x_0)$:
\begin{align}
    \pi_{A}(\bm{x}) & = \pi_0(x_0)\prod_{t=1}^{T} \pi_{t}(x_{t}).
    \label{eq:augdist}
\end{align}
The augmented variable $\bm{x} = (x_0, x_1, \ldots, x_T)$ consists of $T+1$ auxiliary variables. The goal of SMC is to efficiently approximate the entire system $\pi_A(\bm{x})$ with the Monte Carlo approximator, successively sampling $x_t$ in the order $t =0,1, \ldots, T$. The definition of the auxiliary distribution is arbitrary. For example, if the last auxiliary distribution is defined as the original target $\pi_T(x_T) = \pi(x)$, we can take a sample set of $x_T$ to obtain an approximate distribution. Alternatively, all $T$ auxiliary distributions $\pi_1(x_1) \ldots, \pi_T(x_T)$ can be set to be identical to the target $\pi(x)$. In this case, all samples of $x_1, \dots, x_T $ can be used for the approximate inference. The essence of SMC methodology is to be able to use any sequence of auxiliary distributions to efficiently obtain random samples from the intractable target distribution. As a constraint to be satisfied, it is imposed that the last auxiliary distribution is consistent with the target distribution.

To derive the sampling algorithm, we rewrite the augmented distribution in Eq. \ref{eq:augdist} as 
\begin{align}
    \pi_{A}(\bm{x}) = \pi_0(x_0)\prod_{t=1}^{T} \left\{ \frac{\pi_{t}(x_{t})}{\eta(x_{t}|x_{t-1})\pi_{t-1}(x_{t-1})} \right\}\eta(x_{t}|x_{t-1})\pi_{t-1}(x_{t-1}).
    \label{eq:SMCstep}
\end{align}
The conditional probability distribution $\eta(x_t|x_{t-1})$, called the proposal distribution, determines the transition process from $x_{t-1}$ to $x_t$. Assume that we currently have a sample set $\{x_{t-1}^i | i=1, \ldots, m\}$ of $ x_{t-1}$ that follows $\pi_{t-1}(x_{t-1})$. This set defines a Monte Carlo approximation of $\pi_{t-1}(x_{t-1})$ as 
\begin{eqnarray}
\hat{\pi}_{t-1}(x_{t-1}) = \frac{1}{m} \sum_{i=1}^m I(x_{t-1} = x_{t-1}^i) 
\label{eq:expirical_distribution}
\end{eqnarray}
The indicator function $I(\cdot)$ takes the value one if the argument is true; otherwise, it takes zero. The purpose of each step of SMC is to derive a conversion from $\hat{\pi}_{t-1}(x_{t-1})$ to $\hat{\pi}_{t}(x_{t})$ based on the form in Eq. \ref{eq:SMCstep}. To be specific, consider the following recursive formula derived by substituting the approximate distribution $\hat{\pi}_{t-1}(x_{t-1})$ into Eq. \ref{eq:SMCstep}: 
\begin{equation}
\begin{split}
\hat{\pi}_{t}(x_{t}) &= \left\{ \frac{\pi_{t}(x_{t})}{\eta(x_{t}|x_{t-1})\pi_{t-1}(x_{t-1})} \right\}\eta(x_{t}|x_{t-1}) \hat{\pi}_{t-1}(x_{t-1}) \\
&=  \frac{1}{m} \sum_{i=1}^m w(x_t|x_{t-1}^i) \eta(x_{t}|x_{t-1}^i) 
\ \ (t=1, \ldots, T).
\end{split}
\end{equation}
Based on this form with the given $\hat{\pi}_{t-1}(x_{t-1})$, we generate a sample set $\{x_t^{i}| i=1,\ldots, m\}$ of $x_t$ from $\hat{\pi}_{t}(x_{t})$ by performing the sampling importance resampling (SIR) method \cite{rubin1988using} with the importance weight given by $w(x_t | x_{t-1}^i) = \frac{\pi_{t}(x_{t})}{\eta(x_{t}|x_{t-1}^i)\pi_{t-1}(x_{t-1}^i)}$. The algorithm starts with an initial sample set $\{x_0^i| i=1, \ldots, m\}\sim \pi_0(x_0)$ and repeats the following steps for $t=1, \ldots, T$:
\begin{enumerate}

\item[(1)] Draw a particle $x_t^{i}$ from the proposal distribution $\eta(x_t |x_{t-1}^i)$ for each of $i=1, \ldots, m$.

\item[(2)] Calculate the importance weight $w(x_t^i|x_{t-1}^{i})$ to obtain the approximate distribution as
\begin{eqnarray}
\hat{\pi}_{t}(x_t) = \frac{1}{\sum_{i=1}^m w(x_t^{i}|x_{t-1}^{i})}
\sum_{i=1}^m w(x_t^{i}|x_{t-1}^{i}) I(x_t = x_t^{i}).
\end{eqnarray}

\item[(3)] Resample $\{x_t^{i}| i=1,\ldots, m\}$ with probability proportional to $ w(x_t^{i})$ to renew the $m$ samples as following an empirical distribution $\hat{\pi}_{t}(x_t) = \frac{1}{m} \sum_{i=1}^m I(x_t = x_t^i)$ with the equal weight as in Eq. \ref{eq:expirical_distribution}.

\end{enumerate}
In step 1, the previous sample set is tentatively replaced with a new one according to the proposal distribution $\eta(x_t|x_{t-1}^{i})$. The proposal distribution is designed to search a neighboring area of the currently obtained $x_{t-1}^{i}$ with a certain probability and to search a still unexplored region with the remaining probability. This strikes a balance between exploitation and exploration. In step 2, we calculate the importance weight representing the goodness-of-fit of the proposed $x_{t}^{i}$. Finally, in step 3, we determine the survival or death of $x_{t}^{i}$ according to the importance weight. The SMC is essentially equivalent to a genetic algorithm.

As shown later, the simple SMC performs very poorly in our molecular design task. The support of the posterior distribution is extremely large; moreover, the promising solution sets are widely scattered in the huge support. Because conventional SMC cannot cope with the target task, we developed an extended method following the SMC framework.

\subsection*{Recurrent design algorithm for synthetic reaction networks}

\begin{figure}[h!]
\includegraphics[width=0.95\textwidth]{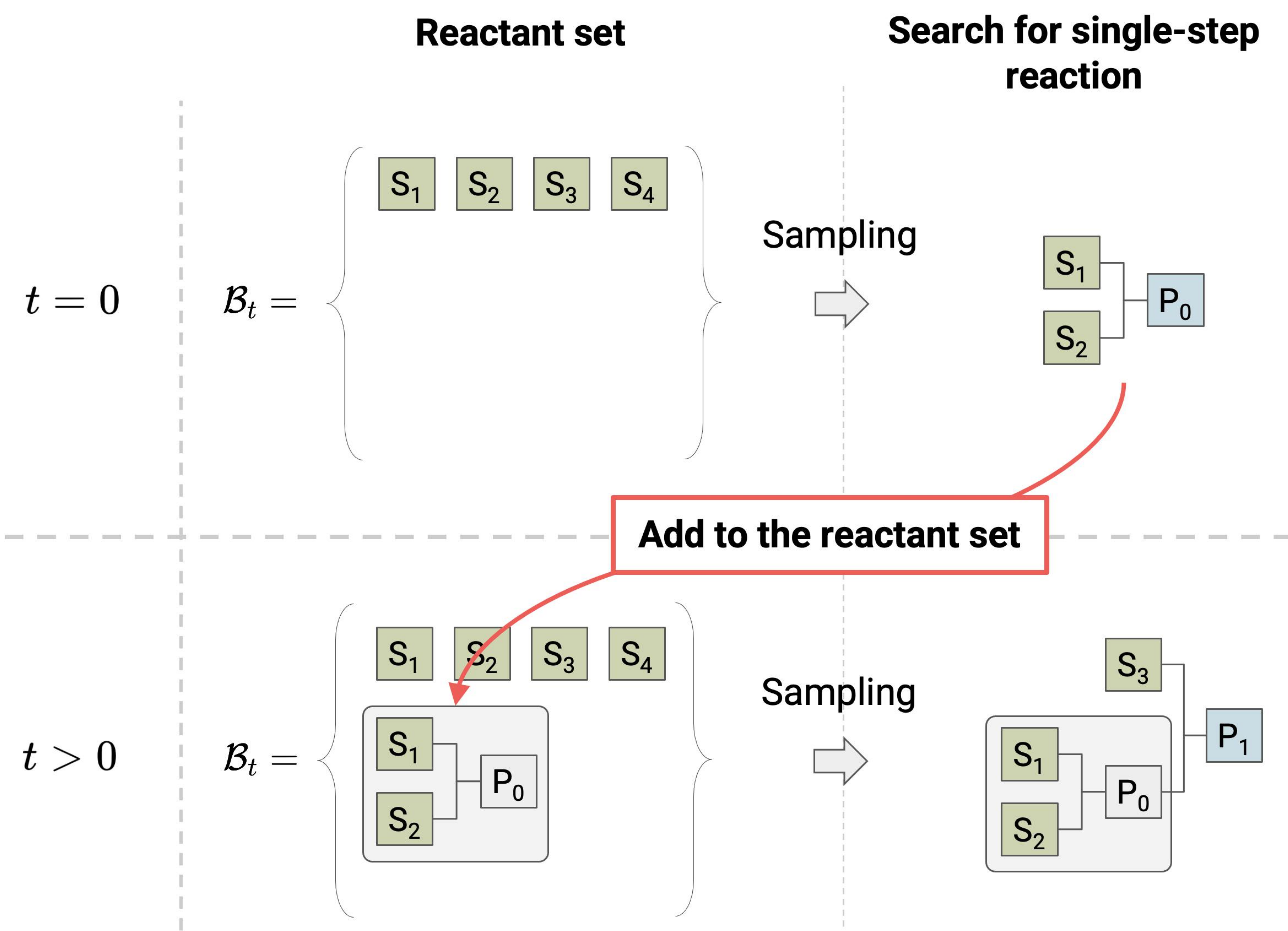}
\caption{
Design of synthetic reaction networks using the SMC calculation based on sequential stacking algorithm. In the current step, all sampled reaction networks and their products are added to the reactant pool in the next step. By searching for single-step reactions from this expanding pool of reactants, the network can be built up recursively.
\label{fig:pool_enrich}
}
\end{figure}

Here, we begin by depicting the key idea of the recurrent design technique for reaction networks following the example shown in Figure \ref{fig:pool_enrich}. Note that for any reaction network, the final product at the root node has two reactants in its parent nodes. We consider that at each step of SMC, denoted by $t$, only the two reactants of the final product are explored. In the forward cascade model $Y= h \circ g(S|G)$, only the single-step reaction $S_1 + S_2 \rightarrow P$ is considered as $G$, and the two reactants $\{S_1, S_2\}$ are selected from the original set $\mathcal{B}$ of commercial compounds and an additional set $\mathcal{I}_{t-1}$ containing all intermediate products computationally synthesized until step $t-1$. Let $\mathcal{B}_t = \mathcal{B} \cup \mathcal{I}_{t-1}$ be the expanded set of candidate reactants and let $x_t=\{S_1, S_2\}$ be the auxiliary variable at $t$. Then, the auxiliary distribution $\pi_t(x_t)$ is defined as 
\begin{eqnarray}
\pi_t(x_t) \propto \exp\left( - \frac{1}{\sigma} d(Y(x_t), U^*) \right) \ \text{where} \ \forall x_t \in \mathcal{B}_t \times \mathcal{B}_t: \ Y(x_t)= h \circ r(S).
\end{eqnarray}
Unlike the general form in Eq. \ref{eq:prop}, the forward model $Y(x_t)= h\circ r(S)$ represents only the single-step reaction model $r(S)$, and consequently the auxiliary distribution $\pi_t(x_t)$ is a function of two reactants only. Nevertheless, the model is able to represent general reaction networks by selecting the already calculated intermediate products contained in $\mathcal{B}_t$, to which their reaction networks are implicitly assigned.

Specifically, we proceed with the following SMC procedure. In step $t-1$, a sample set $\{x_{t-1}^i | i=1,\ldots, m\}$ of reactant set $x_{t-1}$ is obtained, and a new set $\{x_{t}^i | i=1,\ldots, m\}$ is generated using the proposal $\eta(x_t|x_{t-1})$, where each $x_{t}$ is sampled from $\mathcal{B}_t$. Here, all newly generated products $P_i = r(x_{t}^i)$ ($i=1, \ldots, m$) are added to the set of candidate reactants in the next step, as follows:
\begin{eqnarray}
\mathcal{B}_{t+1} = \mathcal{B}_{t} \cup \{P_1, \ldots, P_m\}.
\end{eqnarray}
Thus, the support $\mathcal{B}_t$ of the auxiliary distribution constitutes a sequence of increasing sets as
\begin{eqnarray}
\mathcal{B} = \mathcal{B}_{0} \subseteq \mathcal{B}_{1} \subseteq \cdots \subseteq \mathcal{B}_{T-1} \subseteq \mathcal{B}_{T}.
\end{eqnarray}
At each step, the reactants are sampled from the auxiliary distribution $\pi_t(x_t)$ defined on $\mathcal{B}_{t} \times \mathcal{B}_{t}$ following the procedure described above. After calculating the predicted products $P_i = r(x_{t}^i)$ using the single-step reaction model and predicted properties $Y_i = h \circ r(x_{t}^i)$, and calculating the importance weights, resampling is performed to determine the survival or death of $x_{t}^i $. Because the solution space for each $t$ is restricted to the support of two reactants $\mathcal{B}_{t}\times \mathcal{B}_{t}$, no combinatorial explosion occurs. Additionally, as $t$ increases, the network topology that can be represented by $\mathcal{B}_{t}$ increases monotonically. In principle, if $t$ approaches infinity under the use of a proper proposal distribution, then $\mathcal{B}_t$ can represent any network topology. We call this method the Bayesian sequential stacking algorithm, which describes the process of constructing a reaction network by recurrently stacking single-step reactions.

For the proposal $\eta(x_t| x_{t-1})$, we employ a mixture model such that a neighboring reactant of each in $x_{t-1}$ is selected from $\mathcal{B}_t$ with probability $\alpha$, and with probability $1-\alpha$, $x_{t}$ is chosen completely at random from $\mathcal{B}_t \times \mathcal{B}_t$ in order to obtain a renewed $x_t$. Probability $\alpha$ is a hyperparameter that controls the trade-off between ``exploitation'' and ``exploration''. The ``exploration'' creates a mechanism that enhances the diversity of solutions. The problem we face here is the computational cost of the neighborhood search. The set of candidate reactants $\mathcal{B}_t$ grows monotonically with each step. Calculating the similarity between all entries in $\{x_{t-1}^i | i=1,\ldots,m\}$ and the monotonically increasing $\mathcal{B}_t$ at every step is quite time-consuming. Therefore, we introduce a method to reduce the computational complexity, as described below.

\begin{figure}[h!]
\includegraphics[width=0.95\textwidth]{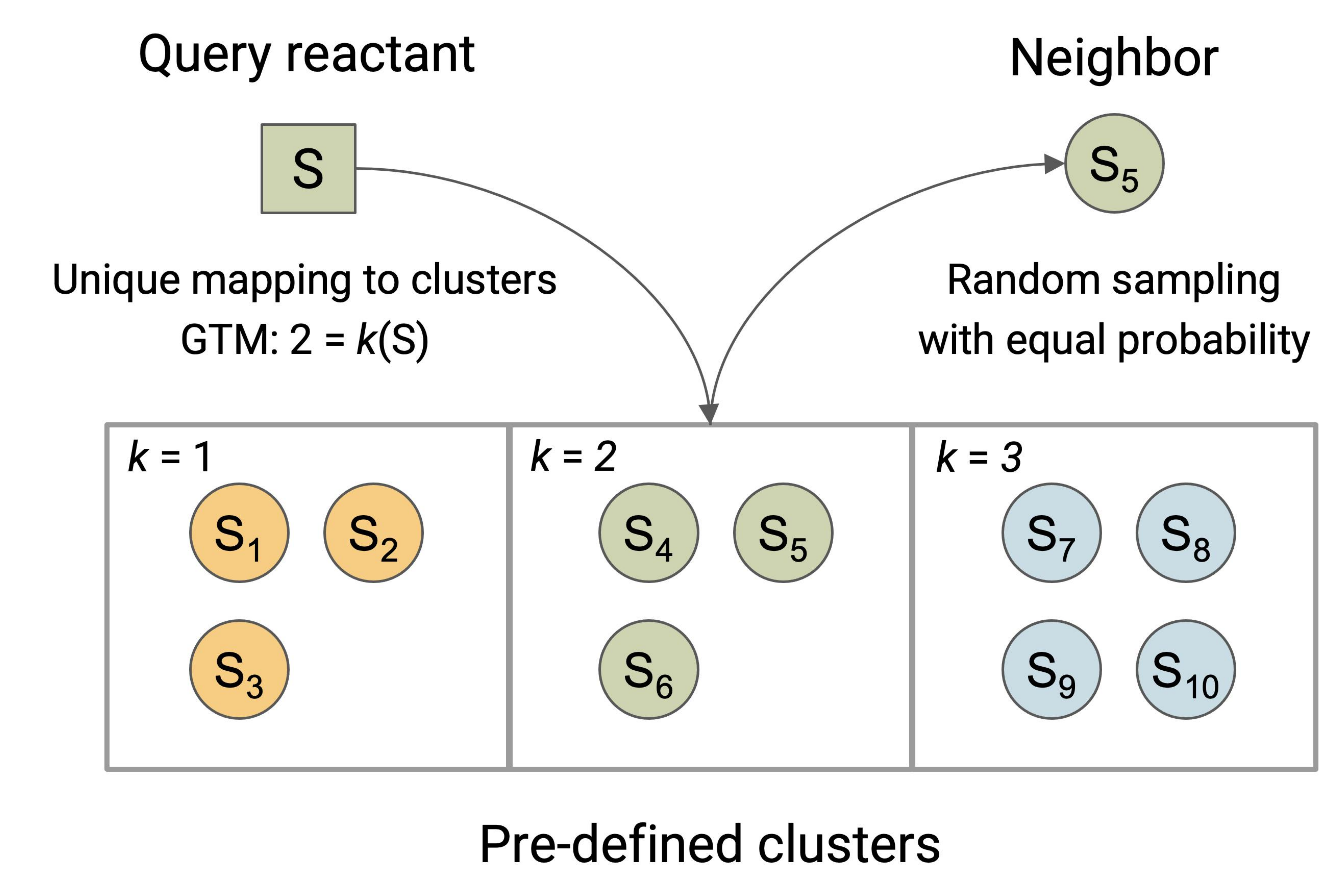}
\caption{Efficient sampling scheme to generate structurally similar reactants based on pre-partitioning of the reactant space. Using a pretrained GTM, any query reactant is uniquely mapped to a cell or cluster to which similar reactants belong. A structurally similar reactant of the query compound can be obtained by sampling the instances in the cell with equal probabilities.}
\label{fig:gtm}
\end{figure}

The initial set $\mathcal{B}$ of commercial compounds is divided into $K$ clusters according to the pattern of the chemical structures. First, the chemical structure $S$ is transformed into a descriptor vector $\phi(S)$ of length 3239 with the concatenation of RDKit fingerprint \cite{landrum2006rdkit} (length 2048), MACCS (Molecular ACCess System) keys \cite{durant2002reoptimization} (length 167), and Morgan fingerprint \cite{morgan1965generation} of radius 2 (length 1024). Then, generative topographic mapping (GTM) is applied \cite{bishop1998gtm}, which is a well-established integrated method of dimensionality reduction and clustering, in order to obtain a unique mapping from vectorized chemical structures to class labels as
\begin{eqnarray}
k = k(S) \equiv k(\phi(S)) \ \text{where} \ k \in \{1, \ldots, K\}.
\end{eqnarray}
Thus, an arbitrary compound set can be partitioned into $K$ disjoint clusters (Figure \ref{fig:gtm}).

Using a trained GTM, candidate reactants that are newly added to $\mathcal{B}_{t-1}$ are sequentially grouped into each of the $K$ predefined clusters $\mathcal{C}_k^{t-1}$ ($k=1, \ldots, K$). Here, $R \in \mathcal{C}_k$ denotes an element of cluster $k$, where the cluster members vary with step $t$; however, we omit the subscript indicating the dependence of the cluster on $t$. The model $\eta_{0}(x_t | x_{t-1})$ for the neighborhood search in the proposal distribution selects $x_{t} = \{S_{1,t}, S_{2,t}\}$ with equal probability from cluster $R \in \mathcal{C}_{k (S_{i,t-1})}$ ($i=1,2$) to which each reactant $S_{1,t-1}$ or $S_{2, t-1}$ in $x_{t-1}$ belongs. The explicit form of the probabilistic model can be expressed as 
\begin{eqnarray}
\eta_{0}(x_t | x_{t-1}) = \prod_{i=1}^2 \prod_{k=1}^K \left( \frac{1}{|\mathcal{C}_k|} \right)^{I(S_{i,t} \in \mathcal{C}_{ k (S_{i, t-1})})}.
\end{eqnarray}
For the model corresponding to ``exploration’’, we sample a candidate from $\mathcal{B}_t$ with equal probability. In summary, the proposal distribution is given by the two-component mixture distribution as
\begin{eqnarray}
\eta(x_t|x_{t-1}) = \alpha \ \eta_0(x_t|x_{t-1}) + (1-\alpha) \prod_{i=1}^2 \left( \frac{1}{|\mathcal{B}_t|} \right)^{I(S_{i, t} \in \mathcal{B}_{t})}.
\end{eqnarray}
The first and second terms function as the ``exploitation’’ and ``exploration’’ mechanisms, respectively. 

\subsection*{Acceleration techniques}

Here, we consider the problem of slow computation of the deep neural network for the reaction prediction. In this study, we used Molecular Transformer developed by Schwaller et al. \cite{schwaller2019molecular} for the forward prediction of a single-step reaction. In our test, a single-step reaction prediction of $m=500$ particles using Molecular Transformer required 25--40 s on average on a Linux server with an NVIDIA V100 GPU. When the number of steps was set as $T=5000$, the total runtime of the reaction prediction exceeded 56 hours.

We therefore used a computationally light surrogate model $Y = d(S)$ that directly predicts property $Y$ from a given set of reactants $S$ without going through the reaction prediction to generate the product. As described later, the chemical structure of each reactant in $S$ was converted into a binary vector of length 3239 by concatenating the Morgan fingerprint with radius 2 and bit length 1024, the MACCS keys, and the RDKit fingerprint. The intersection of the binary vectors for the two reactions in $S$ was then taken to obtain a single-descriptor vector. From the United States Patent and Trademark Office (USPTO) open chemical reaction dataset \cite{lowe2017chemical}, which contains 1.1M reactions, we selected a subset of 492k reactions that involve exactly two reactants. We then substituted the selected reactant pairs in Molecular Transformer to obtain their products and calculated their model properties $Y$ using the property predictor $h$. The gradient boosting regressor \cite{schapire2013boosting} was then trained to learn the mapping from $S$ to $Y$ using 80\% of the 492k instances as the training set. For hyperparameter tuning, using 20 candidate values for the learning rate, 15 for the max depth, and 10 for the number of estimators, we performed 10-fold cross-validation looped within the training set and selected the optimized hyperparameters attaining the smallest mean absolute error (MAE). In the calculation of the importance weights in the SMC module, the surrogate $Y(x_t) = d(S)$ was used instead of $Y(x_t) = h \circ r(S)$. Samples drawn from the proposal distribution at each step of the SMC were handed over to the validation module with Molecular Transformer $r$ and property predictor $h$ for the exact calculation of their synthetic products and resulting properties.

\begin{figure}[h!]
\includegraphics[width=0.95\textwidth]{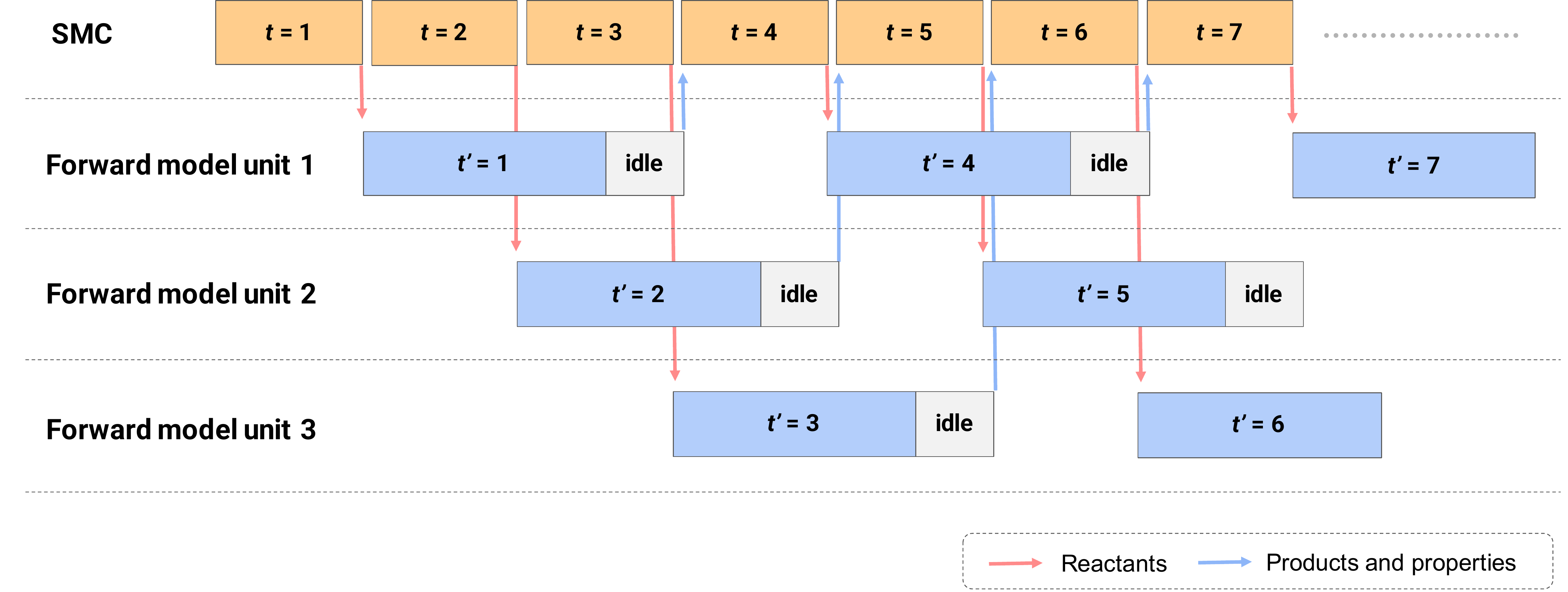}
\caption{
SMC algorithm using asynchronous parallel computation.}
\label{fig:algorithm}
\end{figure}

We also considered a technique for parallel computing, wherein the entire algorithm was divided into two modules: SMC and the aforementioned validation calculation of products and properties using the forward models. The SMC module sequentially feeds the reactants sampled at each step into the module of the forward model. The forward-prediction module calculates the synthetic products and properties of the received reactants and adds the calculated products to the pool of candidate reactants in the SMC module. These two units carry out the computation of the reactant sets flowing from the different steps of SMC. The two modules perform their tasks simultaneously once the data exchange is complete, synchronizing the timing of the data exchange, as schematically described in Figure \ref{fig:algorithm}. The SMC module proceeds with its calculation as $t=0,1,\ldots$ without any suspension (top row). In contrast, the forward-prediction module is subject to an idle state. Once the calculation of one step of the SMC is completed, the set of reactants produced there is handed over to the module for the forward calculation in an idle state. The forward-prediction module then calculates the predicted products using Molecular Transformer, which are sent back to the SMC module. Once the data transfer is completed, the forward-prediction module returns to the idle state and waits to receive the next reactant set. In the example shown in Figure \ref{fig:algorithm}, we assume that the SMC module is approximately 2.5 times faster than the forward-prediction module. In this case, three forward-prediction modules are allocated to different processing units in order to reduce the idling time as much as possible.

\begin{algorithm}[h]
\label{algorithm1}
\begin{flushleft}
\textbf{Initialize} commercially available compounds $\mathcal{B} = \{S_1,...S_N\}$\\
\textbf{Initialize} intermediate products $\mathcal{I}_0=\{\}$\\
\textbf{Initialize} auxiliary distribution $\pi_{0}(x_0),\pi_{1}(x_1),...,\pi_{T}(x_T)$,\\
\textbf{Initialize} number of SMC steps $T$\\
\textbf{Initialize} number of particles $m$\\
\textbf{Initialize} surrogate forward-prediction model $d$\\
\textbf{Do in multiprocess}\\
\vspace{3mm}
\textbf{SMC} \hspace{6cm} \textbf{Forward prediction}
\end{flushleft}
  \caption{De novo design of functional molecules and their synthetic routes}
  \begin{multicols}{2}
    \begin{algorithmic}[1]
      \scriptsize
      \FOR{$t \gets 1$ to $T$}
      \STATE $\mathcal{B}_t\gets \mathcal{B}\cup\mathcal{I}_{t'}$
      \STATE $X_t = \{\}$
      \FOR{$i \gets 1$ to $m$}
      \STATE Draw $x_{t}^{i}$ from $\eta(x_{t}|x_{t-1}^{i})$
      \STATE $X_t \gets X_t \cup \{x_t^i\}$
      \STATE Predict surrogate property $Y_t^i = d(\{S_{1,t}^i,S_{2,t}^i\})$
      \STATE Calculate the importance weight $w(x_t^i)$ by $Y_t^i$
      \STATE Resample $x_{t}^{i} $ from $ \pi_{t}(x_{t})$
      \ENDFOR
      \STATE Pass $X_t$ to the forward-prediction module.
      \ENDFOR
    \end{algorithmic}
    \columnbreak
    \begin{algorithmic}[1]
      \scriptsize
      \FOR{$t' \gets 1$ to $T$}
      \STATE Wait for $X_{t'}$
      \STATE Assign $X_{t'}$ to a processing unit
      \FOR{$x_{t'}^i$ in $X_{t'}$}
      \STATE Predict product $P_{t'}^i = r(x_{t'}^i)$\\
      \STATE Predict property $Y_{t'}^i = h(P_{t'}^i)$\\
      \ENDFOR
      \STATE $\mathcal{I}_{t'} \gets \mathcal{I}_{t'}\cup \{P_{t'}^i|i=1,\cdots m\}$
      \STATE Pass $\mathcal{I}_{t'}$ to the SMC module.
      \ENDFOR
    \end{algorithmic}
  \end{multicols}
  \begin{flushleft}
  \textbf{Output:} $\{x_t^i|Y_t^i\in U^{*}; t=0,\cdots T; i=1,\cdots m\}$
  \end{flushleft}
  \label{alg:main}
\end{algorithm}

\subsection*{Summary: Bayesian sequential stacking algorithm for molecular design}

The proposed molecular design algorithm is summarized in Algorithm \ref{alg:main}. The overall workflow consists of two modules: SMC and calculation of the forward model. The SMC module performs posterior sampling of the single-step reactions using the current pool $\mathcal{B}_t$ of the candidate reactants. The sampling calculation consists of (1) resampling based on the goodness-of-fit of the current set of reactants, (2) updating the set of hypothetical reactants based on the proposed distribution, and (3) calculating the goodness-of-fit importance weights. All reactant sets generated from the proposal distribution are handed over to the forward model module to generate the predicted products and characterize their properties. All the products generated in this module are successively added to the pool of candidate reactants $\mathcal{B}_t$ in the SMC module. The monotonically growing reactant pool $\mathcal{B}_t$ contains the intermediate products calculated from the reaction prediction model. Therefore, throughout the search for single-step reactions in SMC, reaction networks with various structures can be constructed by sampling and stacking intermediate products for which their reaction networks have already been calculated. The computational complexity of the neighborhood search owing to the monotonic growth of $\mathcal{B}_t$ is suppressed by pre-clustering the set of reactants using the GTM. In summary, the algorithm is based on four ideas: (1) the recurrent algorithm for the network search, which is based on the sequential expansion of the reactant pool, (2) the avoidance of the neighborhood search from large reactant pools using GTM clustering, (3) efficient computation of the forward cascade model using a surrogate model, and (4) the asynchronous parallel computation algorithm.

\section*{Results}
The predictive and computational performances of the proposed method were evaluated using an application example that involved designing drug-like molecules. In particular, we constructed several variants of the Bayesian molecular design algorithms by combining the four constituent mechanisms described in the previous section, and compared their performance to quantitatively investigate their individual contributions to the overall scheme.

\subsection*{Target properties}
As target properties $Y=h(P)$, we considered the following two physicochemical properties that quantitatively express the drug-likeliness of a designed molecule $P$. 
\begin{itemize}
    \item \textbf{QED} The quantitative estimate of drug-likeness (QED) quantifies drug-likeness as a score ranging between 0 and 1 \cite{bickerton2012quantifying}. QED was modeled on a dataset of 771 known oral drugs using eight descriptors, including molecular weight, polar surface area, and number of hydrogen bond donors and acceptors. The higher the QED value, the more drug-like the molecule is judged to be. The target range of the QED was set to be greater than 0.8.
    \item \textbf{logP} The octanol-water partition coefficient logP is defined as the normal logarithm of the ratio of the concentrations of molecules in the organic and aqueous layers at equilibrium. According to Lipinski's rule of five \cite{lipinski1997experimental}, the target range of logP was defined as not exceeding 5.
\end{itemize}
Computational models of these two properties are implemented in the modules of RDKit\cite{landrum2006rdkit}.

\subsection*{Reaction prediction}
As a forward response prediction model $P = g(S)$, we employed Molecular Transformer \cite{schwaller2019molecular}, which is known to be a state-of-the-art model. This attention-based neural translation model defines a translation between the SMILES strings of reactants and their products. For simplicity, reagents were removed from the model input. SMILES strings for multiple reactants were input into the model as separated by periods ``.’’. The SMILES strings were all canonicalized using RDKit. The inputs were tokenized with the regular expression according to the original paper of Molecular Transformer. 

The model was trained from scratch using 80\% of the training instances randomly selected from the 500k recorded reactants and products in the USPTO dataset \cite{lowe2012extraction}. The top 1 prediction accuracy of the trained model for the remaining data reached 78.2\%, which is comparable to the accuracy reported in previous studies \cite{guo2020bayesian, pesciullesi2020transfer}.

\subsection*{Surrogate models}

We applied gradient boosting regression to construct surrogate models that predict the two target properties from a set of two reactants without going through the prediction of synthetic products using Molecular Transformer. We randomly selected two pairs of 492K reactants from USPTO, generated their products using Molecular Transformer, and calculated their QED and logP. The resulting set of 394K samples was used to train the surrogate models for QED and logP. The $R^2$ values for predicting QED and logP for the 98K additional test samples were 0.004 and 0.172, respectively.

\subsection*{Commercial compounds} 

We used a subset of the Enamine building block catalog global stock as the set of commercially available reactants by which virtual molecules are synthesized \cite{enamine}. This set, consisting of 150K unique building blocks, was narrowed down by Gottipati et al. \cite{gottipati2020learning}. to those applicable to one or more rules in a template-based reaction prediction model. The design task was to identify the synthesizable products in this reactant set and to meet the required properties. The Enamine reactant set was also used to train the GTM model as well. Each reactant was transformed into a binary vector of length 3239 concatenated with the Morgan fingerprint of radius 2 and bit length 1024, the MACCS key, and the RDKit fingerprint. A total of 441 predefined clusters were used in the similarity-based SMC proposal.

\subsection*{Sequential Monte Carlo} 
As explained previously, the proposed method was designed based on four ideas. To investigate the contribution of each idea to the overall performance, we implemented the five variants listed in Table \ref{tab:methods}: (1) SMC-RECUR-GTM-SR-PL, (2) SMC-RECUR-GTM, (3) SMC, (4) Random, and (5) Random-RECUR. The meanings of the abbreviations attached to each method are as follows:
\begin{itemize}
    \item \textbf{RECUR} Recurrent design technique for the reaction networks
    \item \textbf{GTM} The use of the 441 GTM clusters in the similarity-based proposal
    \item \textbf{SR} Surrogate models for the direct prediction of QED and logP
    \item \textbf{PR} Asynchronous parallel computing of the SMC and reaction prediction modules
\end{itemize}
For the vanilla SMC in (3), only single-chain synthetic reactions with the number of steps fixed at $n \in \{1, 2, 3\}$ were considered. For example, in the case of $n = 2$, we considered only the cascade-type reactions as $S_{11} + S_{12}\rightarrow P_1, P_{1} + S_{21}\rightarrow P_2$ and searched for three reactants $S = \{ S_{11}, S_{12}, S_{21}\}$ that synthesize product $P=P_2$ satisfying the property requirements. For ``Random’’ in (4), a completely random search for two reactants was conducted for the single-step reaction $S_{11} + S_{12} \rightarrow P_1$. The variant ``Random-RECUR’’ in (5) represents an integrated method of random search and recursive network design algorithm.

The experimental conditions were set to be common to the five variants: for SMC, the number of iterations was set to $T=10000$, the number of particles was set to $m=500$, and the exploration-exploitation trade-off parameter of the proposed distribution was set to $\alpha = 0.8$, indicating a 20\% chance of making an exploratory search.

\begin{table}[]
    \centering
    \caption{Five different algorithms to be evaluated for the performance test. The abbreviations are explained in the main text.}
    \begin{tabular}{ p{4cm} p{1cm} p{4.5cm} p{1cm} }
         \hline
         Method & SMC & Network type & Parallel \\
        \hline
         SMC-RECUR-GTM & \checkmark & any & -\\
         SMC-RECUR-GTM-SR-PL & \checkmark & any & \checkmark \\
         SMC-n    & \checkmark & $n$-step single chain ($n \in \{1,2,3\}$) &  - \\
         Random    & -  &  one-step single chain      & - \\
         Random-RECUR & -  & any  & -\\
         \hline
    \end{tabular}
    \label{tab:methods}
\end{table}

\subsection*{Computational environments}
The experiments were carried out on an NVIDIA DGX STATION with 4 Tesla V100 and Intel Xeon E5-2698 v4 CPUs. 

\subsection*{Results and discussion}

After the experiment, we compared the five methods listed in Table \ref{tab:methods}. As mentioned above, for the vanilla SMC, the design of the single-chain reaction routes was tested under three different step sizes $n \in \{1,2,3\}$. Therefore, seven methods were included in this comparison. Each method was tested in two scenarios. In scenario 1, a relatively small number of commercial compounds was assumed to be available: the candidate reactants were generated by randomly sampling 10k of the 150k Enamine compounds available for purchase. In scenario 2, all 150k compounds were used to design the molecules.

\begin{figure}[h!]
\includegraphics[width=0.95\linewidth]{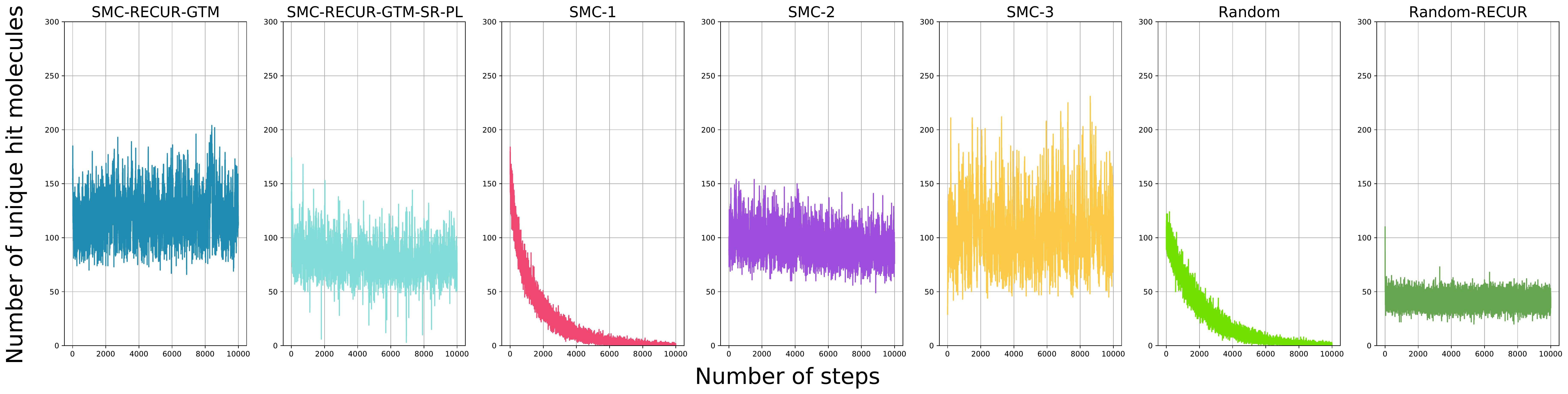}
\caption{Number of design molecules reaching the target property range at each step for scenario 1 where the number of commercially available reactants is limited to 10K. The correspondence between the seven methods and their labels is shown in Table \ref{tab:methods}.}
\label{Fig:efficiency_per_step}
\end{figure}

First, we report the results of scenario 1. As a performance measure, we simply considered the number of unique designed molecules that reached the target property range. Figure \ref{Fig:efficiency_per_step} shows the evolution of the number of unique hit molecules at each step of the sequential calculation. Unsurprisingly, the two random search algorithms were clearly less efficient than the others. For SMC-1 and Random, whose design space is restricted to single-step reactions with two reactants, the number of hits decayed as the number of search steps increased. This means that the limited design space consisting of two combinations of at-most 10k commercial compounds can be adequately covered by the naive algorithms. In contrast, for the vanilla SMC with two- and three-step reactions (SMC-2, SMC-3) and the recurrent algorithms to explore arbitrary reaction networks (SMC-RECUR-GTM, SMC-RECUR-GTM-SR-PL), the number of hit molecules at each step remained constant, and no decay trend was observed for $T=10000$. This observation confirms that the search space for general reaction design is quite large. In the comparison between SMC-RECUR-GTM and its extended version, SMC-RECUR-GTM-SR-PL with the surrogate models shows that the number of hits of the latter is slightly lower than that of the former. This is a natural consequence because the surrogate models involve approximation errors.

\begin{figure}[h!]
\includegraphics[width=0.95\linewidth]{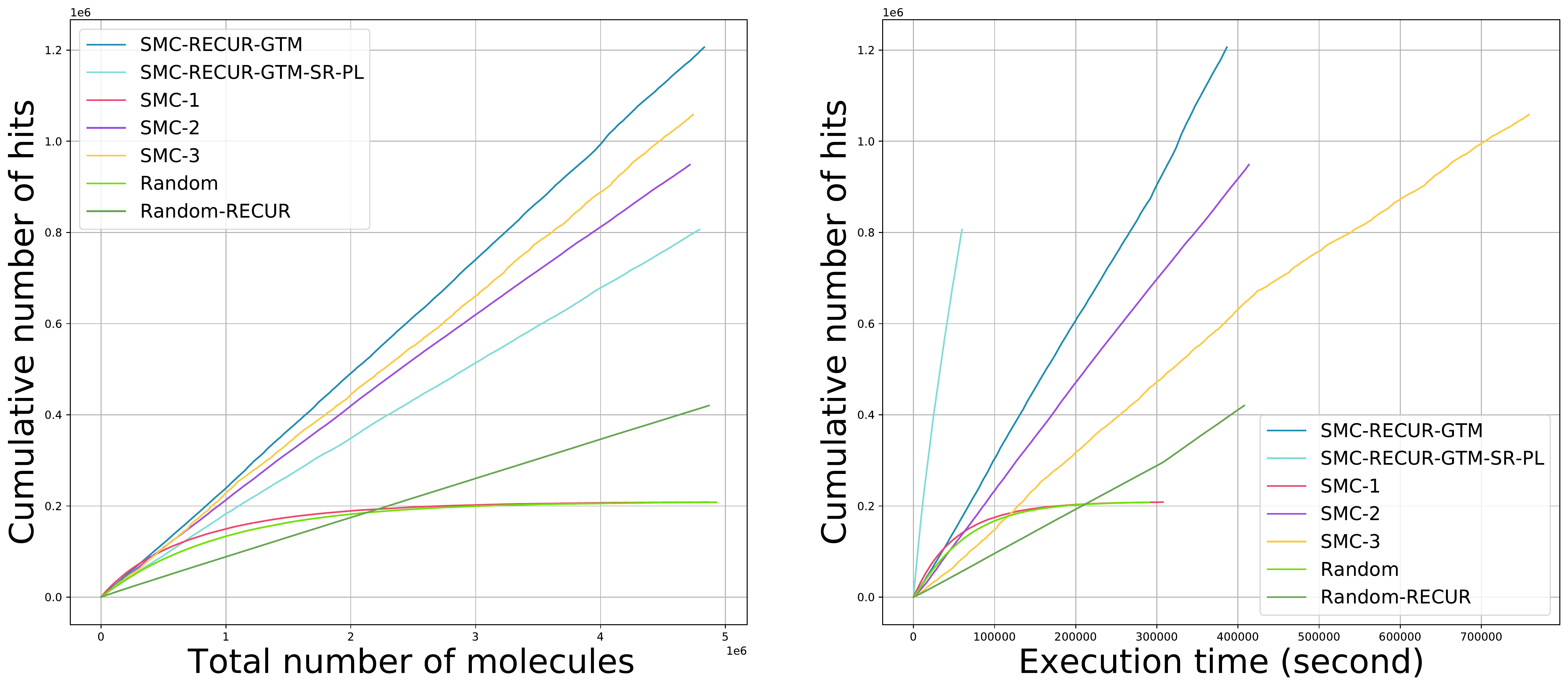}
\caption{Cumulative number of hit molecules in scenario 1 where the number of commercially available reactants is limited to 10K. The left and right plots show the cumulative number of hits as a function of the total number of molecules generated and the execution time (CPU time), respectively.}
\label{Fig:number_of_promising}
\end{figure}

Figure \ref{Fig:number_of_promising} shows the evolution of the cumulative number of unique hit molecules that reached the target region against the number of steps. The left panel shows the cumulative number of hits as a function of the total number of molecules generated. This is a different view of the results in Figure \ref{Fig:efficiency_per_step}. It was confirmed that SMC-RECUR-GTM is the most efficient method to find molecules hidden in the target region. It was also confirmed that SMC-RECUR-GTM-SR-PL has a reduced hit rate due to the use of the surrogate models involving approximation errors, as mentioned above. Vanilla SMC-2 and SMC-3 also consistently found the target molecules. This suggests that it is difficult to fully cover the large design space with as few as $T \times m =500000$ search trials. The right panel in Figure \ref{Fig:number_of_promising} shows the cumulative number of hits per step as a function of execution time (right panel). In terms of computational cost, the parallel recursive molecular design algorithm SMC-RECUR-GTM-SR-PL with the surrogate models showed by far the highest search efficiency.

\begin{figure}[h!]
\includegraphics[width=0.95\linewidth]{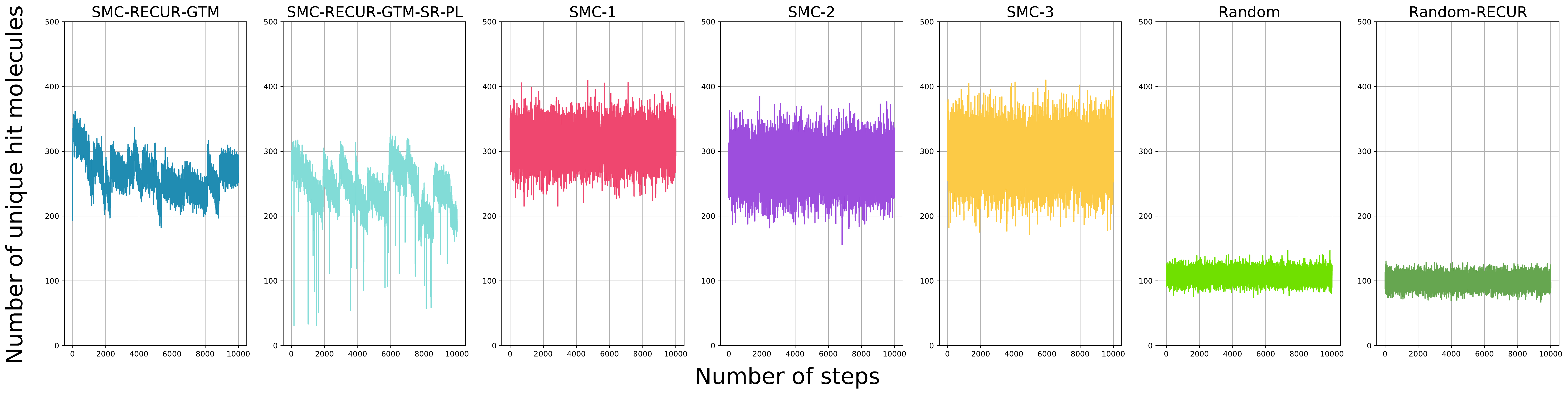}
\caption{Number of design molecules reaching the target property range at each step for scenario 2 where all 150K commercial compounds were used in the design. }
\label{Fig:efficiency_per_step_big}
\end{figure}

\begin{figure}[h!]
\includegraphics[width=0.95\linewidth]{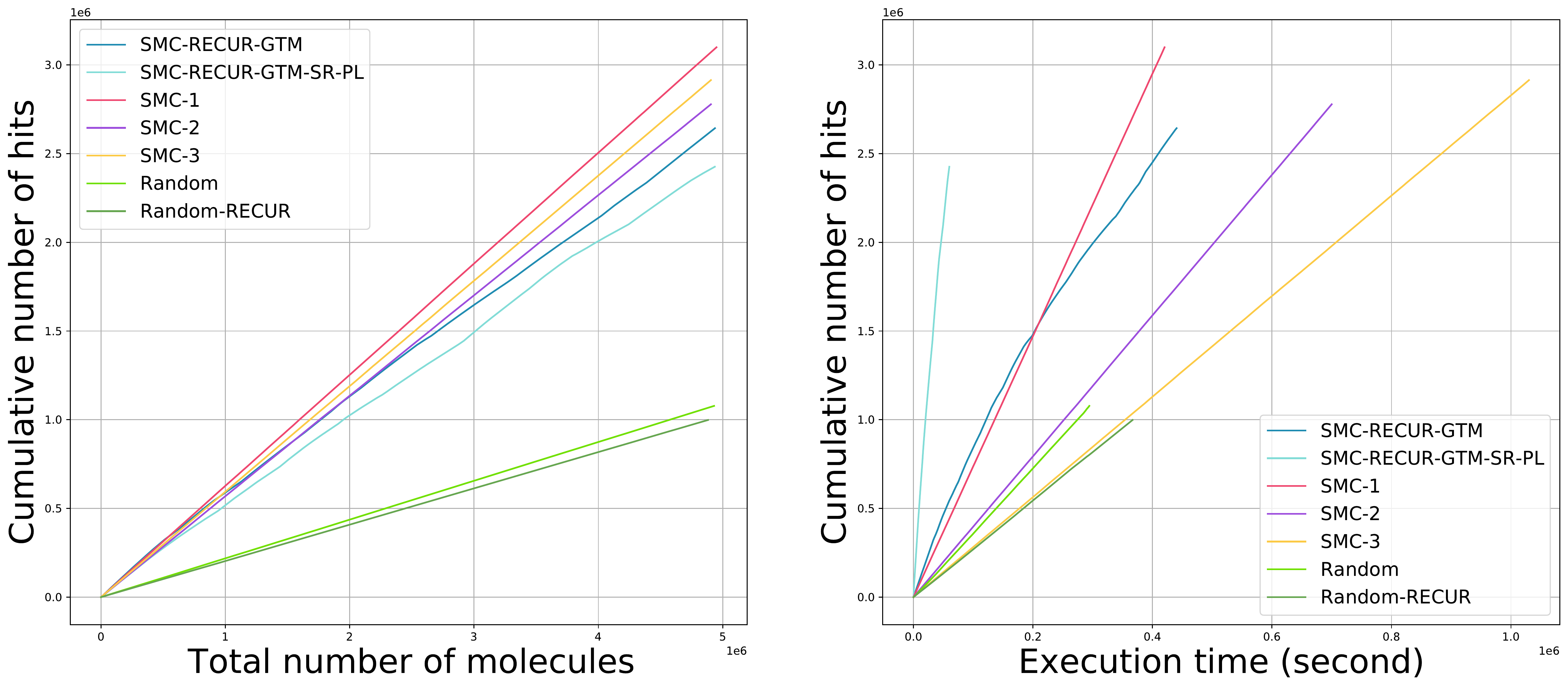}
\caption{Cumulative number of hit molecules in scenario 2 where all 150K commercial compounds were used in the design. The left and right plots show the cumulative number of hits as a function of the total number of molecules generated and the execution time (CPU time), respectively.}
\label{Fig:number_of_promising_big}
\end{figure}

With regard to the results for scenario 2: The number of hit molecules at each step (Figure \ref{Fig:efficiency_per_step_big}) and the cumulative number of hits (Figure \ref{Fig:number_of_promising_big}) were significantly lower for the two random searches, as in scenario 1. In contrast to the results of scenario 1, SMC-1 and Random, whose design space was restricted to single-step reactions with two reactants, did not show any decay in the number of hit molecules throughout all steps. This observation indicates that, as the number of commercial reactants increases, the design space becomes much larger, even for single-step reactions. Among the methods other than the random searches, no significant difference was found in the evolution of the number of hit molecules and the cumulative number. However, similar to the results of scenario 2, SMC-RECUR-GTM-SR-PL showed by far the highest search efficiency in terms of the number of hit molecules per execution time (right panel in Figure \ref{Fig:number_of_promising_big}). From a practical point of view, search performance against execution time is considered the most important criterion. Therefore, we conclude that SMC-RECUR-GTM-SR-PL is superior in terms of the evaluation criteria for detecting the number of unique molecules.

\begin{figure}[h!]
\includegraphics[width=0.95\linewidth]{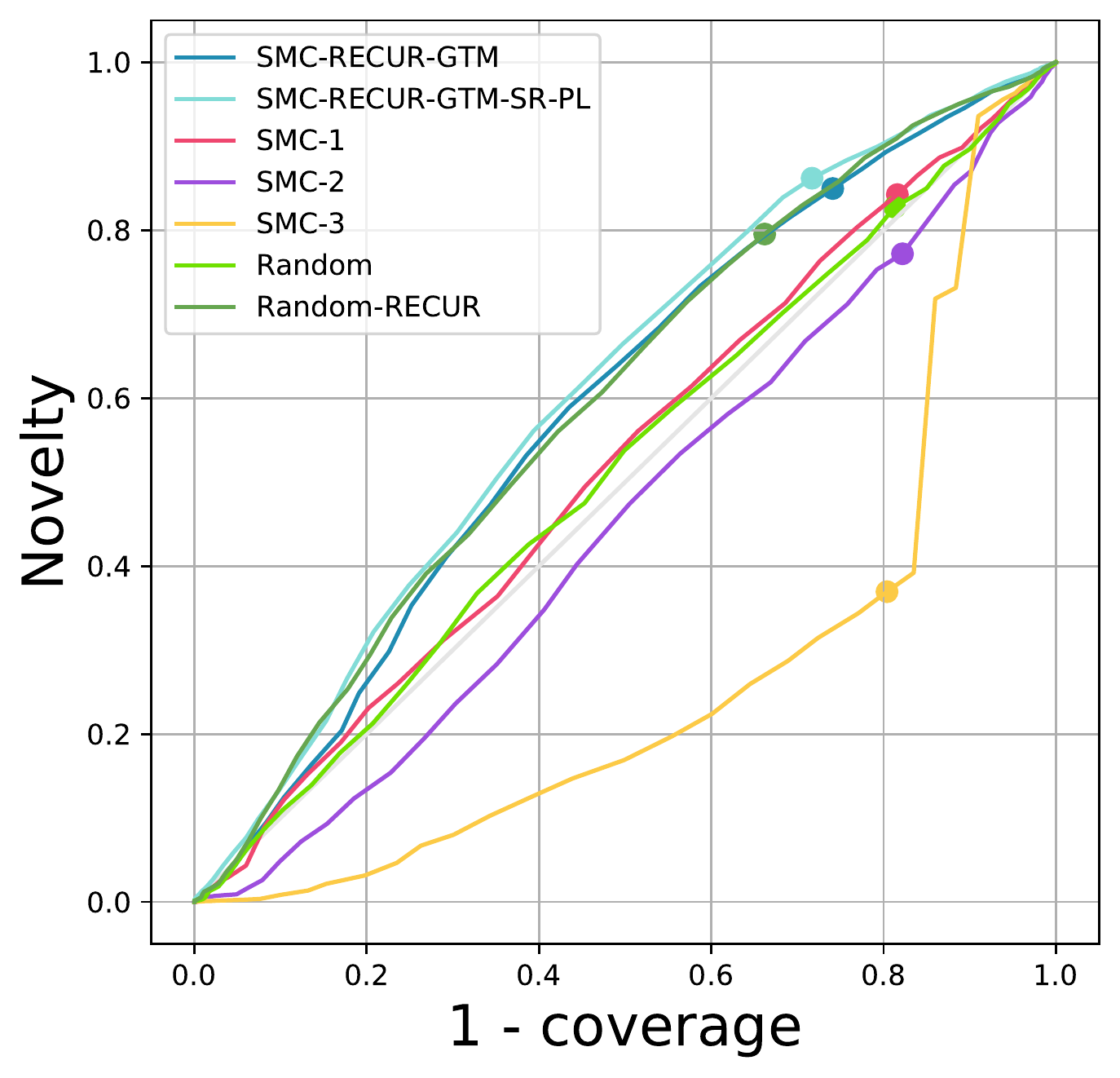}
\caption{Coverage (horizontal axis) and novelty (vertical axis) of the set of designed virtual molecules with respect to existing molecules in ChMBLE, which is drawn as a function of varying thresholds of Tanimoto similarity. The circle represents the coverage and novelty when the similarity threshold is set to 0.7.}
\label{Fig:ChEMBL_roc}
\end{figure}

We then performed a quality assessment based on the structural diversity and novelty of the generated hit molecules and their coverage against existing molecules synthesized so far. Specifically, we assessed the coverage and novelty of the 300K hypothetical molecules that reached the target region in scenario 2 against a hit compound set extracted from the organic compound database ChEMBL \cite{gaulton2017chembl} as follows:
\begin{enumerate}
\item[(1)] The set of hit compounds, denoted by $A$, was obtained by calculating the two properties of 127k compounds registered in the ChEMBL.
\item[(2)] Let $B$ be the set of 300K hit virtual molecules generated in scenario 2.

\item[(3)] Evaluate the similarity between the compounds in $A$ and $B$ using the Tanimoto coefficient of the ECFP fingerprint descriptor (radius 2, bit length 2048).
\item[(4)] Set the threshold values of the Tanimoto coefficient as $\gamma \in \{0, 0.1, ..., 1\}$.
\item[(5)] Coverage: Calculate the percentage of molecules in $A$ with a similarity greater than $\gamma$ to those in $B$.

\item[(6)]
Novelty: Calculate the percentage of molecules in $B$ with a similarity smaller than $\gamma$ to those in $A$.

\item[(7)]
Vary the threshold $\gamma$ from 0 to 1, and draw a curve representing the balance between coverage and novelty (CN curve), as in Figure \ref{Fig:ChEMBL_roc}.
\end{enumerate}
The CN curve shows an upward or downward convex pattern depending on the inclusive relationship of the distributions of $A$ and $B$.
Ideally, a set of reasonably novel and diverse molecules should be generated, while maintaining a high coverage to existing molecules. This corresponds to a situation in which the distribution of $B$ encompasses $A$. In such a case, the CN curve deviates slightly from the 45$^{\circ}$ line and shows an upward convex pattern. For this criterion, SMC-RECUR-GTM, SMC-RECUR-GTM-SR-PL, and Random-RECUR showed better properties than the others (Figure \ref{Fig:ChEMBL_roc}). When the threshold of Tanimoto similarity was set to $\gamma \ge 0.7$, the coverage and novelty of SMC-RECUR-GTM and SMC-RECUR-GTM-SR-PL were both approximately 0.3 and 0.8, respectively.

\begin{figure}[h!]
\includegraphics[width=0.95\linewidth]{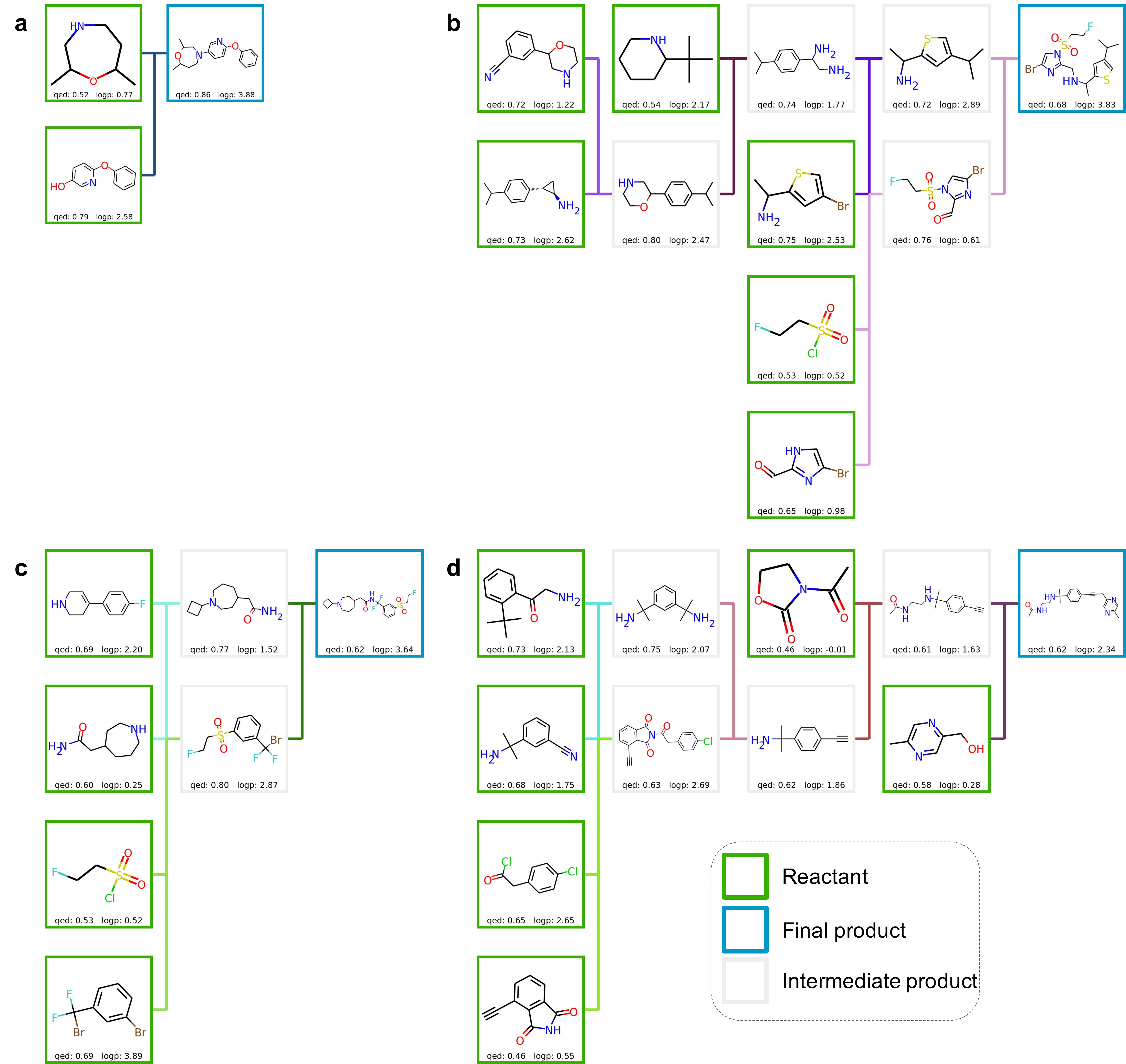}
\caption{Four examples of designed molecules and their synthetic reaction networks, which are drawn with the Python code we have made available online at \cite{ssr}. Single-step reactions are depicted by different colors. Ten design examples are provided in Supporting Information (Additional file 1: Section S1).}
\label{Fig:route_examples}
\end{figure}

The analyses presented here can be performed using the Python package Seq-Stack-Reaction, which we have made available online under the license BSD 3-Clause \cite{bsd3}. Seq-Stack-Reaction \cite{ssr} can be installed with Conda \cite{anaconda}. Users can plug-in any model for property and synthetic reaction predictions. The list of commercial compounds and the maximum number of reaction steps can also be specified arbitrarily. In a parallel computing environment, the asynchronous parallel search algorithm can be executed by specifying an option. Additionally, a visualization of the predicted synthetic pathway network of the excavated products is implemented. Figure \ref{Fig:route_examples} shows some examples of the visualization for the four identified products and their synthetic pathways. In Supporting Information (Additional file 1: Section S1), 10 predicted products and their synthetic pathway networks are shown as examples.

In the future, we need to develop methods and platforms to systematically evaluate the properties of molecules designed by machine learning as well as the practicality of the designed synthetic pathways. The computation of molecular property characterization and synthetic reactions using quantum chemical methods is very time intensive; therefore, the implementation of a high-throughput evaluation system is quite challenging. Of course, evaluation by exhaustive experiments is also very costly and cannot be a realistic solution. The establishment of a systematic evaluation method is an important issue for future research.

\section*{Conclusions}

In this study, we developed a machine learning methodology and a general-purpose Python library for simultaneously designing molecules exhibiting any desired set of properties and their synthetic reactions with any network topology. A forward model was constructed using a synthetic reaction prediction model and property prediction models as building blocks, and its inverse mapping was obtained based on a Bayesian inference framework to simultaneously identify a reaction network, its constituent reactant sets, and the final products that satisfy arbitrary target properties. The reactant set was selected based on a combination of predefined commercial compounds. As the design space, consisting of arbitrary reaction networks and reactant sets, was very large, we developed a sequential Monte Carlo algorithm incorporating a recurrent algorithm for the network search. Performance tests showed that our algorithm can successfully find high-quality virtual molecules. The quality of the designed molecules was evaluated based on their reproducibility and novelty with respect to previously synthesized molecules. The distributed Python library was designed using an interface that allows users to plug-in arbitrary reaction prediction models, property prediction models, and a set of commercial compounds. This is expected to facilitate its widespread use in practical applications.

Although machine learning-based research for molecular design and synthetic pathway design has been actively pursued in recent years, most such studies have worked independently on the two subjects so far. Thus, research on the simultaneous design of functional molecules and synthetic pathways has not progressed significantly. In particular, few general-purpose libraries are currently available. The aim of this study was to develop a generic methodology and tools to link these two subjects. We believe that this milestone has been achieved.

\begin{backmatter}
\subsection*{Availability of data and materials}
The data on synthetic reactions and ground-truth compounds are openly available at USPTO \cite{lowe2017chemical} and ChEMBL \cite{gaulton2017chembl}, respectively. The Python open-source library of the present method is available at https://github.com/qi-zh/Seq-Stack-Reaction.

\subsection*{Competing interests}
The authors declare no competing interests.

\subsection*{Funding}
This work was supported in part by MEXT under ``Program for Promoting Researches on the Supercomputer Fugaku’’ (Grant Number JPMXP1020210314), JST CREST (Grant Number JPMJCR19I3), a JSPS Grant-in-Aid for Scientific Research (A) (19H01132) from the Japan Society for the Promotion of Science, and a MEXT KAKENHI Grant-in-Aid for Scientific Research on Innovative Areas (19H05820).

\subsection*{Authors' contributions}
Qi Zhang and Ryo Yoshida designed the research; Qi Zhang and Ryo Yoshida wrote the manuscript; Qi Zhang, Chang Liu, and Stephen Wu developed the Python code and performed the analysis; Ryo Yoshida supervised the research.

\bibliographystyle{bmc-mathphys} 
\bibliography{bmc_article}


\section*{Additional Files}
  \subsection*{Additional file 1 --- Sample additional file title}
    10 predicted products and their synthetic reaction networks are shown as examples.

\end{backmatter}
\end{document}